\documentclass[aps,prl,amsmath,amssymb,superscriptaddress,showpacs,twocolumn]{revtex4-1}

\usepackage{graphicx} \usepackage{color} \usepackage{verbatim} 
\usepackage{hyperref}
\def\K{\mathrm{K}}
\def\eV{\mathrm{eV}}
\def\meV{\mathrm{meV}}

\def\V2O3{V$_2$O$_3$}
\def\NNO{NdNiO$_3$ }
\def\LAO{LaAlO$_3$ }

\begin{document}

\title{ Shining light on transition metal oxides: unveiling the hidden
  Fermi Liquid}

\author{Xiaoyu Deng}
\affiliation{Department of Physics and Astronomy, Rutgers University,
  Piscataway, New Jersey 08854, USA}
\author{Aaron Sternbach} \affiliation{Department of Physics,
  University of California-San Diego, La Jolla, California 92093, USA}

\author{Kristjan Haule}
\affiliation{Department of Physics and Astronomy, Rutgers University,
  Piscataway, New Jersey 08854, USA}

\author{D. N. Basov} \affiliation{Department of Physics,
  University of California-San Diego, La Jolla, California 92093, USA}

\author{Gabriel Kotliar}
\affiliation{Department of Physics and Astronomy, Rutgers University,
  Piscataway, New Jersey 08854, USA}

\date{\today}

\begin{abstract}
  We use low energy optical spectroscopy and first principles LDA+DMFT
  calculations to test the hypothesis that the anomalous transport
  properties of strongly correlated metals originate in the strong
  temperature dependence of their underlying resilient quasiparticles.
  We express the resistivity in terms of an effective plasma frequency
  $\omega_p^*$ and an effective scattering rate $ 1/\tau^*_{tr}$. We
  show that in the archetypal correlated material \V2O3, $\omega_p^*$
  increases with increasing temperature, while the plasma frequency
  from partial sum rule exhibits the opposite trend .  $
  1/\tau^*_{tr}$ has a more pronounced temperature dependence than the
  scattering rate obtained from the extended Drude analysis. The
  theoretical calculations of these quantities are in quantitative
  agreement with experiment.  We conjecture that these are robust
  properties of all strongly correlated metals, and test it by
  carrying out a similar analysis on thin film \NNO on \LAO substrate.
\end{abstract}


\maketitle 

Understanding the transport properties in metallic states of strongly
correlated materials is a long-standing challenge in condensed matter
physics.  Many correlated metals are not canonical Landau Fermi
liquids (LFL) as their resistivities do not follow the $T^2$ law in a
broad temperature range.  Fermi liquid behavior emerges only below a
very low temperature scale, $T_{LFL}$, which can be vanishingly small
or hidden by the onset of some form of long range order.  Above
$T_{LFL}$, the resistivity usually rises smoothly and eventually
exceeds the Mott-Ioffe-Regel limit, entering the so-called ``bad
metal" regime\cite{Badmetal-Emery} with no clear sign of saturation
\cite{Mottlimit-Universality-Hussey,
  ResistivitySaturation-RMP-Gunnarsson}. As stressed in Ref.\,
\onlinecite{Badmetal-Emery} an interpretation of the transport
properties in terms of quasiparticles (QPs) is problematic when the
mean free path is comparable with the de Broglie wavelength of the
carriers and describing the charge transport above $T_{LFL} $ is an
important challenge for the theory of strongly correlated materials.

It was shown in the context of the interacting electron phonon system,
that the QP picture is actually valid in regimes that fall outside the
LFL hypothesis\cite{QPs-KineticEquation-Prange}. There are peaks in
the spectral functions which define renormalized QPs even though the
QP scattering rate is comparable to the QP energy. The transport
properties can be formulated in terms of a transport Boltzman kinetic
equation for the QP distribution function, which has precisely the form
proposed by Landau. Solving the transport equation, the dc
conductivity can be expressed as
\begin{equation}
\sigma_{dc}= (\omega_p^*)^2{\tau^*_{tr}} /{4\pi}
\end{equation}
in analogy with Drude formula. The effective transport scattering rate
$1/\tau_{tr}^*$ characterizes the decay of the current carried by QPs
due to collisions involving umklapp effects, and $\omega_p^*$ is the
low energy effective plasma frequency of QPs and can be expressed in
terms of QP velocities and the Landau
parameters\cite{FL-Pines-Nozieres}.

%The low energy effective plasma frequency, in the spherically symmtric case,
%$(\omega_p^*)^2 = (1+{F_1 \over d}) \sum_k (-\frac{\partial
%  f}{\partial k}) (v_k^*)(v_k^*)$ can be expressed in terms of the QP
%velocities and the Landau parameter $F_1$.  $d$ is the
%dimensionality.
%

The temperature dependence of the transport coefficients beyond the
scope of LFL and many salient features seen in correlated oxides, such
as their low coherence scale, non saturating resistivities and
anomalous transfer of spectral weight, are described well in studies
of doped Hubbard model within the framework of dynamical field mean
theory (DMFT) (for early reviews of this topic see \cite{
  Transport-DMFT-Pruschke, DMFT-RMP-Georges}). A complete understanding
of the transport anomalies has been reached
recently\cite{Model-QP-Palsson, Model-RQP-Deng, Model-HiddenFL-Xu}. As
in the Prange-Kadanoff theory\cite{QPs-KineticEquation-Prange}, the
QPs are resilient surviving in a broad region above $T_{LFL}$
\cite{Model-RQP-Deng} and a quantum kinetic equation provides a
quantitative description of the transport\cite{Model-HiddenFL-Xu} .

While in the electron-phonon coupled system treated in
Ref.\,\onlinecite{QPs-KineticEquation-Prange} the Fermi liquid
parameters such as the quasiparticle velocities and therefore
$\omega_p^*$ are temperature independent, they are strongly
temperature dependent in the doped Mott insulator within DMFT due to
changes in the Fermi surface at high temperatures
\cite{Model-QP-Palsson, Model-RQP-Deng} and a strong temperature
dependence of the effective mass at intermediate temperatures
\cite{Model-RQP-Deng, Model-HiddenFL-Xu}.  This strong temperature
dependence of $\omega_p^*$ hides the more conventional temperature
dependence of $1/\tau^*_{tr}$ in the resistivity, which is quadratic
in a broad region of temperatures and has saturating behavior at high
temperatures \cite{Model-HiddenFL-Xu}.  Strong temperature dependence
in the QP electronic structure with the resulting temperature
dependence of $\omega_p^*$ and $1/\tau^*_{tr}$ thus provides a simple
scenario to describe the anomalous transport of correlated metals.

In this Letter, we provide experimental and theoretical evidences that
this picture holds beyond the DMFT treatment of simplified Hubbard
Model, and is indeed relevant to real materials.  We focus on \V2O3.
This archetypal correlated material provided the first experimental
corroboration of the validity of the DMFT picture of Mott
transition\cite{Optic-DMFT-Rozenberg} and is still a subject of
intense experimental studies\cite{V2O3-CriticalBehavior-Limelette,
  V2O3_microscopic_Lupi, V2O3-XAS-Rodolakis, V2O3-Optics-Stewart,
  V2O3-HPmetal-Ding}. We propose formulas to extract the effective
plasma frequency $\omega_p^*$ and effective scattering rate
$1/\tau_{tr}^*$ from optical conductivity and show that they display
the predicted temperature dependence. We contrast their temperature
dependence to that of the plasma frequency and scattering rate
extracted from the standard extended Drude analysis.

In correlated systems the optical conductivity is usually parametrized
with the so-called extended Drude analysis  in terms of two frequency
dependent quantities, the scattering rate $1/\tau(\omega)$ and the mass
enhancement $m^*(\omega)/m_b$ \cite{Optics-RMP-Basov},
\begin{equation}
\sigma(\omega)=\sigma_1(\omega)+i\sigma_2(\omega)=\frac{\omega_p^2}{4\pi}\frac{1}{-i\omega
 \frac{m^*(\omega)}{m_b} +1/\tau(\omega)}.
\end{equation}
The plasma frequency $\omega_p$ is obtained with the partial sum rule
$\frac{\omega_p^2}{8}=\int_0^\Omega \sigma_1(\omega)d\omega $ and
depends on the cutoff $\Omega$ chosen so as to exclude interband
transitions. To test the theory, instead we focus on quantities that
have a simple QP interpretation, namely $ {1/\tau^*_{tr}}$ and
$(\omega_p^*)^2$, from low frequency optical conductivity extracted as
follows,
\begin{equation}(\omega_p^*)^2=4\pi\frac{\sigma_1^2+\sigma_2^2}{\sigma_2/\omega}|_{\omega\rightarrow
  0}, \qquad
1/\tau_{tr}^*=\frac{\sigma_1}{\sigma_2/\omega}|_{\omega\rightarrow 0},
\label{eq:lowfrequencyDrude}
\end{equation}
When a direct determination of the imaginary part of optical
conductivity ( as for example in ellipsometry measurements) is not
available, they can be extracted from $\sigma_1(\omega)$ only, using
\begin{equation}
\frac{\sigma_2(\omega)}{\omega}|_{\omega\rightarrow
    0}=-\frac{1}{\pi}\int^\infty_{-\infty}\frac{1}{\omega'}\frac{\partial
  \sigma_1(\omega')}{\partial \omega'}d\omega'.
\end{equation}
Comparing with extended Drude analysis, we have
$(\omega_p^*)^2=\frac{m_b}{m^*(0)}\omega_p^2$,
$\frac{1}{\tau_{tr}^*}=\frac{m_b}{m^*(0)}\frac{1}{\tau(0)}$. Thus this
analysis is related to extended Drude analysis, but free of partial
sum rule. Similar low frequency analysis has been used in previous
works\cite{CaRuO3-Drude-Kamal, Drude-SimpleMetal-Youn,
  opticreview-Millis, LectureNotes-Armitage, Optics-RMP-Basov},
however the temperature dependence of $\omega_p^*$ was not the focus
of those studies.

We apply the proposed analysis to \V2O3, a prototype of metal
insulator transition (MIT) \cite{V2O3-MIT-McWhan-1,
  V2O3-MIT-McWhan-2}. Pure \V2O3 is a paramagnetic metal (PM) at
ambient condition. It enters antiferromagnetic insulating state (AFI)
below $T_N\simeq150\K$ with a concomitant structural transition, and
the AFI can be quenched by Ti-doping or pressure. The PM can be turned
into the paramagnetic insulator (PI) by slight Cr-doping, which
induces a first order isostructural transition with a small change in
$c/a$ ratio, indicating a typical band-controlled MIT
scenario\cite{MIT-RMP-Imada}.  This first order transition ends at a
second order critical point at temperature around 400$\K$
\cite{V2O3-MIT-McWhan-2, V2O3-CriticalBehavior-Limelette}. The PM
phase exhibits significant signatures of correlations, for instance, a
pronounced QP peak and a broad lower Hubbard band were revealed in
photoemission spectroscopy measurements \cite{V2O3-PES-Mo,
  V2O3-PES-Rodolakis, V2O3-PES-Fujiwara}. The PM phase is a Fermi
liquid at low temperature when AFI state is suppressed
\cite{V2O3-MIT-McWhan-4}.

\begin{figure}
\includegraphics[width=0.47\columnwidth]{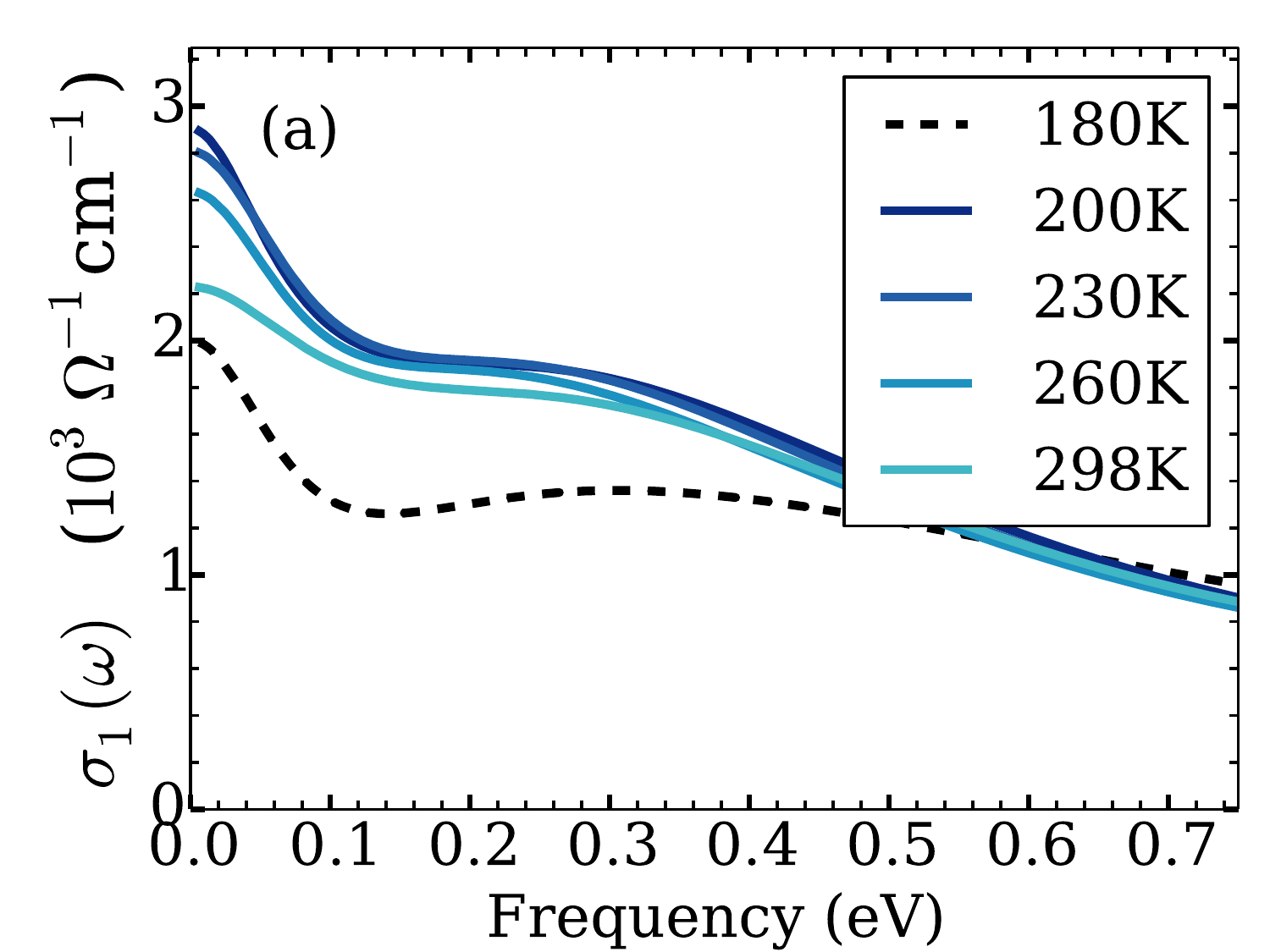}
\includegraphics[width=0.47\columnwidth]{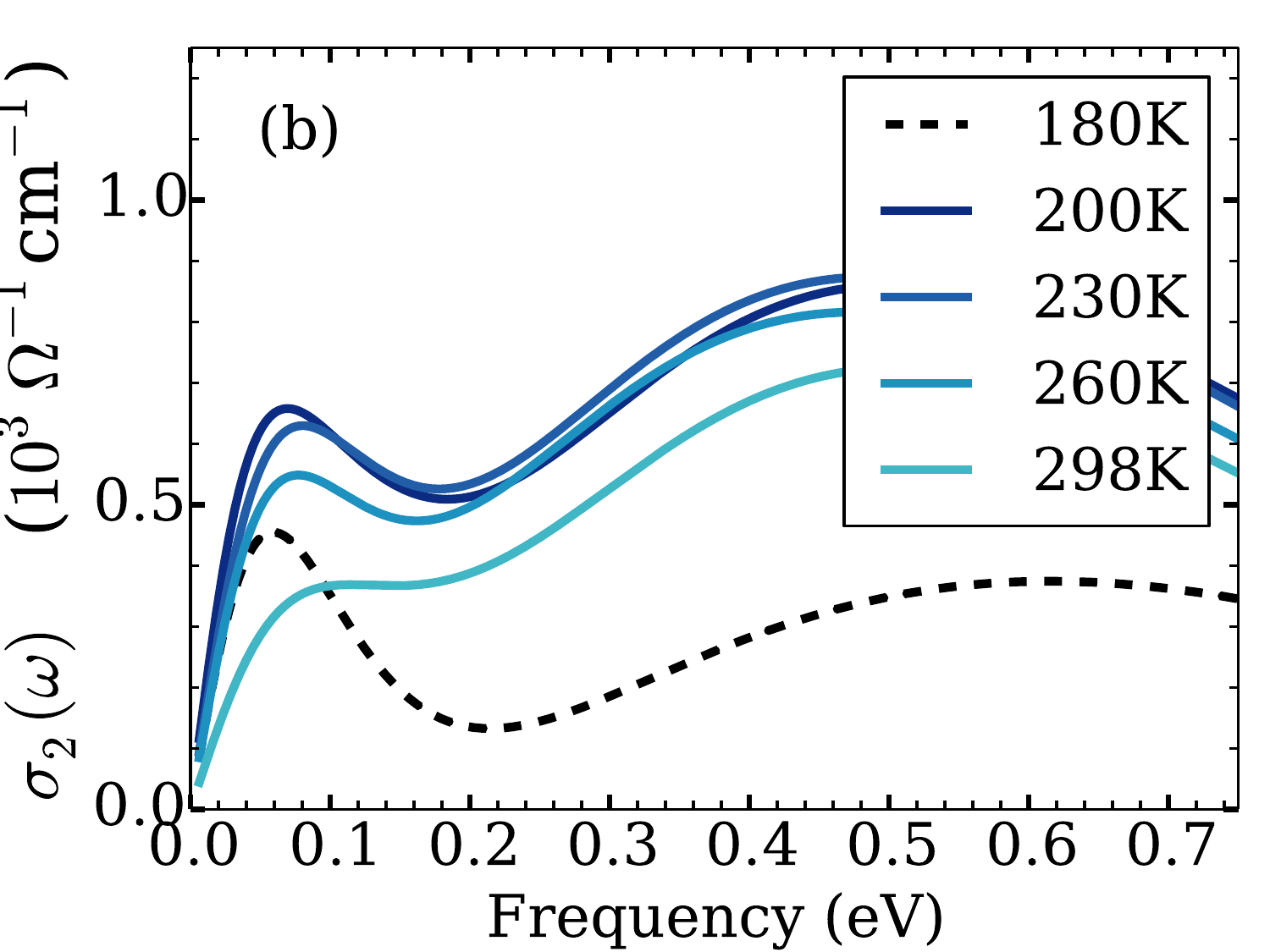}
\includegraphics[width=0.47\columnwidth]{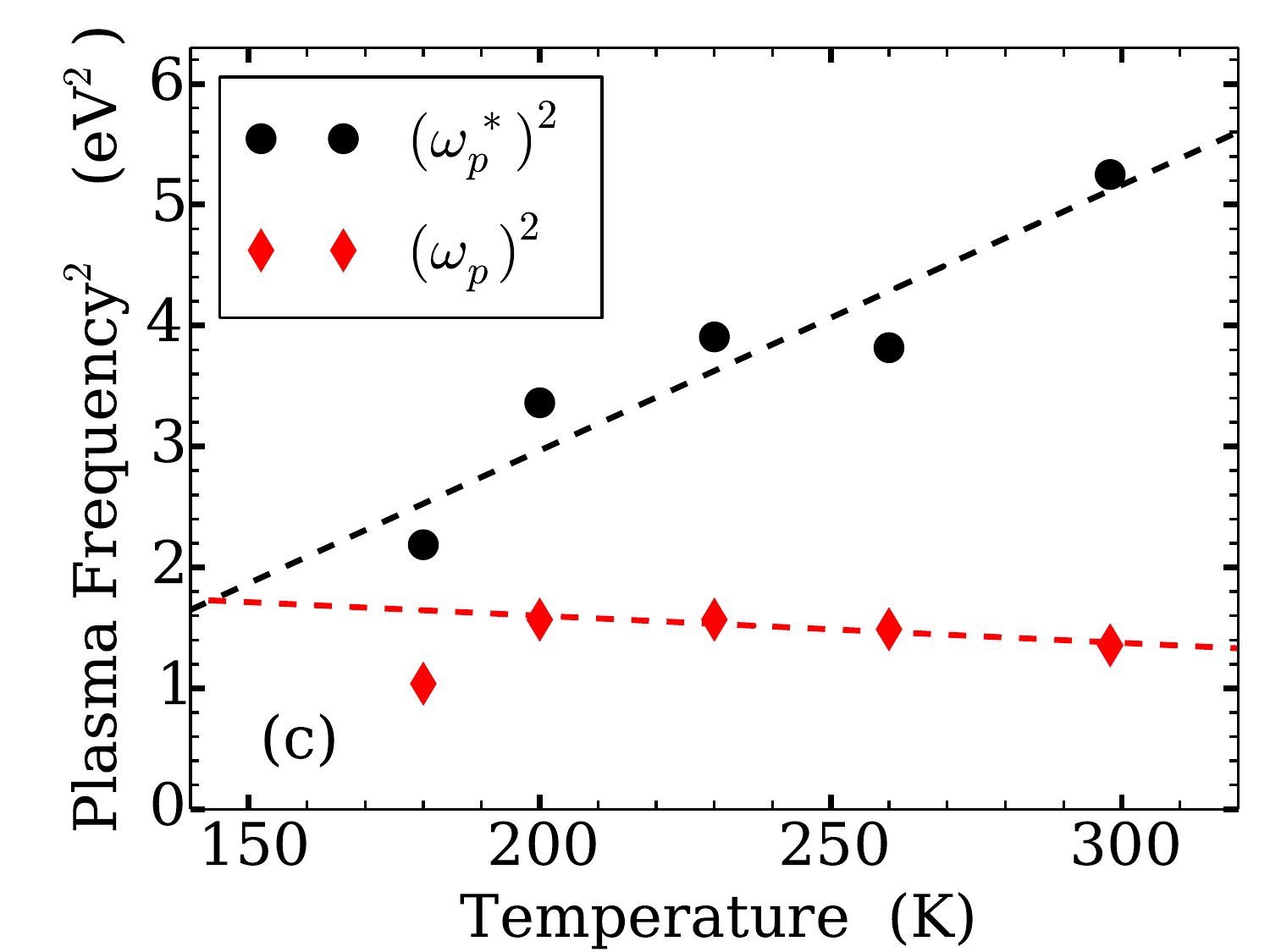}
\includegraphics[width=0.47\columnwidth]{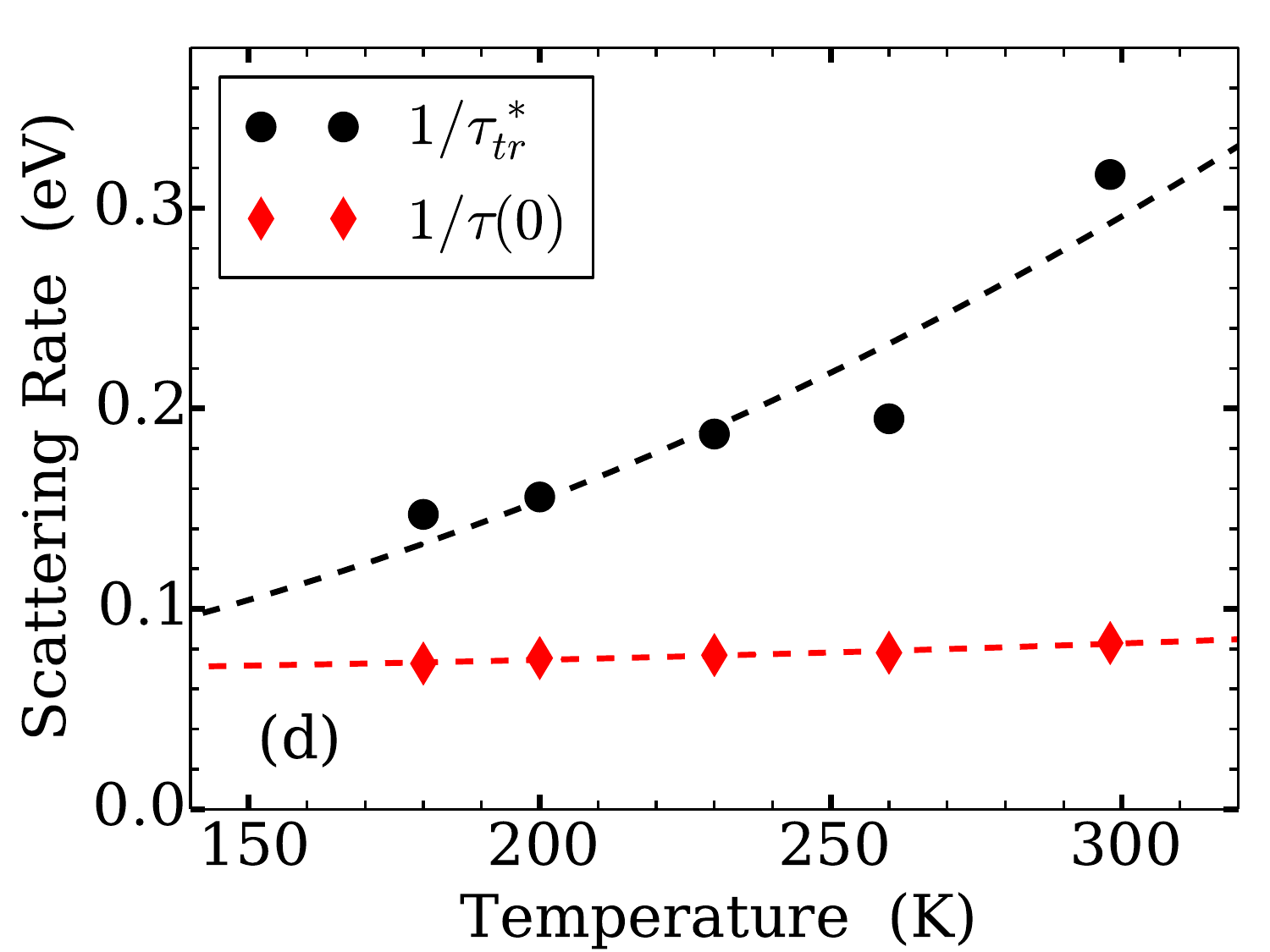}
\caption{Optical conductivity (a) $\sigma_1(\omega)$ and (b)
  $\sigma_2(\omega)$ of \V2O3 at different temperature is taken from
  Ref.~\cite{V2O3-Optics-Stewart}, where dashed lines indicate data at
  $T=180\K$ very close to MIT. (c) $(\omega^*_p)^2$ and (d)
  $1/\tau_{tr}^*$ of \V2O3 are extracted according to
  Eqn.~\ref{eq:lowfrequencyDrude}. For comparison, $\omega_p^2$ and
  $1/\tau(0)$ extracted from the extended Drude analysis are also
  shown. Dashed lines are guides for the eyes by fitting
  $(\omega_p^*)^2$ ($\omega_p^2$, excluding $T=180\K$) and
  $1/\tau_{tr}^*$ ($1/\tau(0)$) to linear and parabolic functions
  respectively.}
\label{wsq_V2O3} 
\end{figure}

Fig~\ref{wsq_V2O3}(a)(b) shows the measured optical conductivity
$\sigma(\omega)=\sigma_1(\omega)+i\sigma_2(\omega)$ of pure \V2O3 in
PM phase\cite{V2O3-Optics-Stewart}. Pronounced Drude peaks show up
even when the resistivity is high (of the order of
$1\;m\Omega^{-1}cm^{-1}$) and does not follow
$T^2$-law\cite{V2O3-MIT-McWhan-3, V2O3-CriticalBehavior-Limelette}.
The Drude peak diminishes gradually upon increasing temperature,
except at the lowest temperature where the transport is probably
affected by the precursor of ordered phase.  $\omega_p^*$ and
$1/\tau^*_{tr}$ extracted according to Eqn.~\ref{eq:lowfrequencyDrude}
are shown in Fig~\ref{wsq_V2O3}(c)(d).  We find that $(\omega_p^*)^2$
increases with increasing temperature. This is in contrast with the
Drude plasma frequency square $(\omega_p)^2$ obtained by the partial
sum rule with a cut off $\Omega=140\meV$, which slightly
decreases\cite{V2O3-Optics-Stewart} except at the lowest
temperature. $1/\tau^*_{tr}$ increases with increasing temperature and
has the same trend as the scattering rate extracted with extended
Drude analysis at zero frequency $1/\tau(0)$, but with a much stronger
temperature dependence.  The experimental data is consistent with an
$(\omega_p^*)^2$ which has a term linear and a $1/\tau_{tr}^*$ which
is quadratic in temperature, revealing a Fermi liquid behavior that is
hidden in $ 1/\tau_{tr}^* $.  The analysis of the experimental data,
thus corroborates the main qualitative predictions of the DMFT
description of transport properties in simple model
Hamiltonian\cite{Model-HiddenFL-Xu}.

We now argue that realistic LDA+DMFT \cite{LDADMFT-RMP-Gabi,
  LDADMFT-review-Held} calculations describe well the optical
properties as well as the extracted quantities $\omega_p^*$ and
$1/\tau^*_{tr}$, hence a local approximation, which ignores vertex
corrections, is sufficiently accurate to capture the experimental
trends.  LDA+DMFT investigations on \V2O3 by several groups have
successfully described the properties of this material near the MIT
\cite{V2O3-DMFT-Held, V2O3-DMFT-Laad, V2O3-DMFT-Poteryaev,
  V2O3-review-Hansmann, V2O3-DMFT-Grieger}.  The correlation in \V2O3
is due to the partially filled narrow $d$-orbitals with a nominal
occupancy $n_d=2$.  The two electrons mainly populate the $e_g^\pi$
and $a_{1g}$ states of vanadium due to surrounding oxygen octahedron
with trigonal distortion.

We perform the LDA+DMFT calculations with an implementation as
described in Ref.~{\cite{LDADMFT_Haule}}, which is based on WIEN2k
package{\cite{wien2k}}. We use projectors within a large (20$\eV$)
energy window constructing local orbitals. This study thus includes
explicitly the oxygen orbitals hybridizing with the d orbitals. The
interaction is applied to $e_g^\pi$ and $a_{1g}$ orbitals only. To
solve the impurity problem, we use continuous-time quantum Monte-Carlo
method with hybridization expansion~\cite{ 
  CTQMC_Haule, CTQMC_Werner}. The Brillouin zone integration is performed with a
regular 12x12x12 mesh. The muffin-tin radius is 1.95 and 1.73 Bohr
radius for V and O respectively. The structure is taken from pure \V2O3 at room
temperature and only paramagnetic state is considered.

The Coulomb interaction $U$ and the Hund's coupling $J$ are set to
$6.0\eV$ and $0.8\eV$ respectively. They are consistent with the ones
used in previous studies \cite{V2O3-DMFT-Held, V2O3-DMFT-Laad,
  V2O3-DMFT-Poteryaev, V2O3-review-Hansmann, V2O3-DMFT-Grieger}, but
$U$ is slightly larger. This is because in previous studies the
relevant correlated orbitals are more extended due to downfolding or
projection onto a small energy window and hence experience a reduced
repulsion. With these parameters the calculated total spectra is
consistent with experiment photoemission spectroscopy measurements
\cite{V2O3-PES-Mo, V2O3-PES-Rodolakis, V2O3-PES-Fujiwara}. The
occupancies of $e_g^\pi$ and $a_{1g}$ orbitals at $T=200K$ are 1.60
and 0.50 respectively, in good agreement with X-ray absorption
spectroscopy \cite{V2O3-XAS-Park} measurements and previous LDA+DMFT
calculations \cite{V2O3-review-Hansmann}. These parameters place \V2O3
on the metallic side but close to the MIT in the phase diagram (
temperature versus interaction strength) with the temperature at
critical endpoint of the first order MIT close to the experimental
findings \cite{V2O3-MIT-McWhan-2, V2O3-CriticalBehavior-Limelette}.

\begin{figure}
\includegraphics[width=0.75\columnwidth]{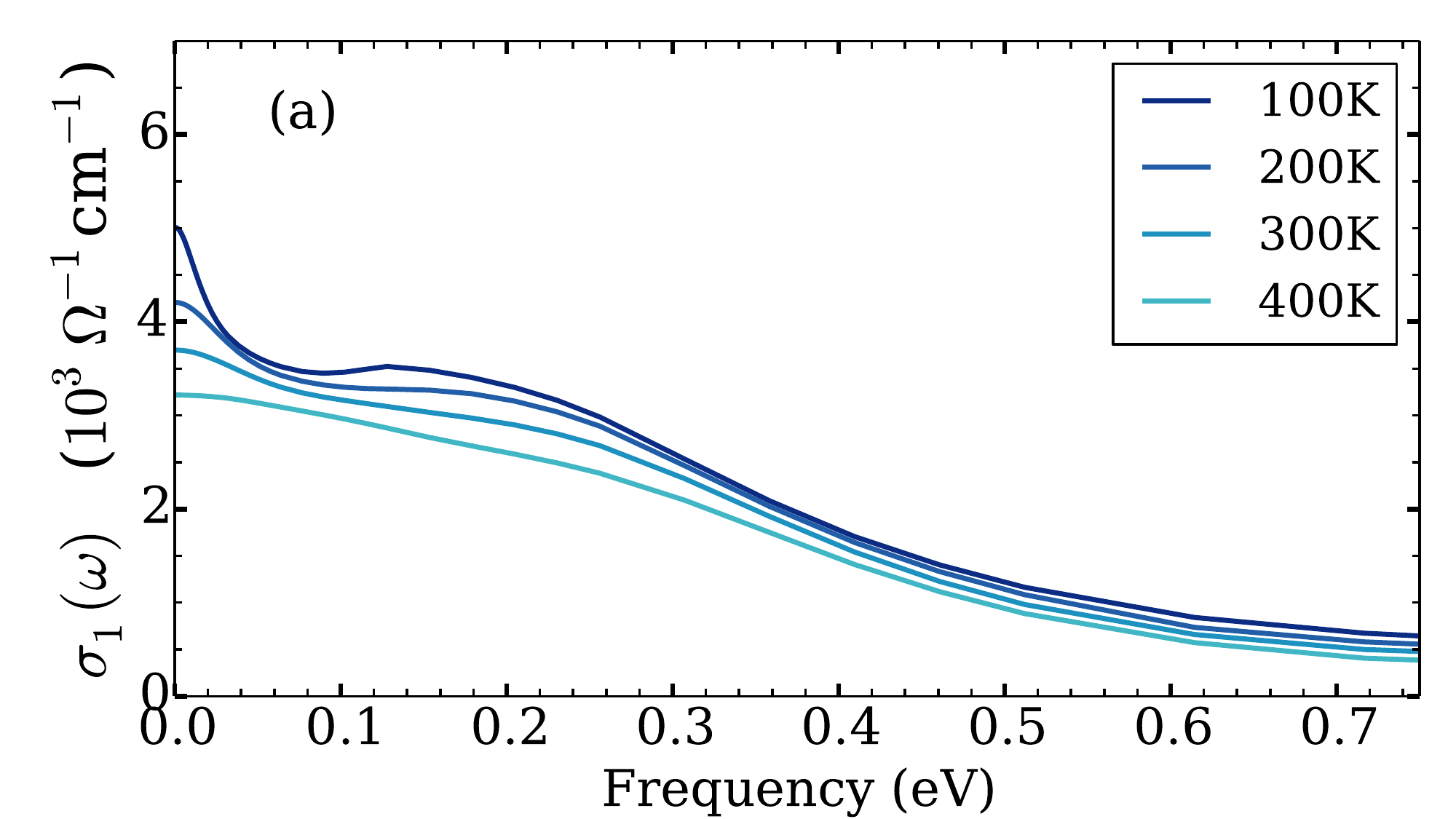}
\includegraphics[width=0.47\columnwidth]{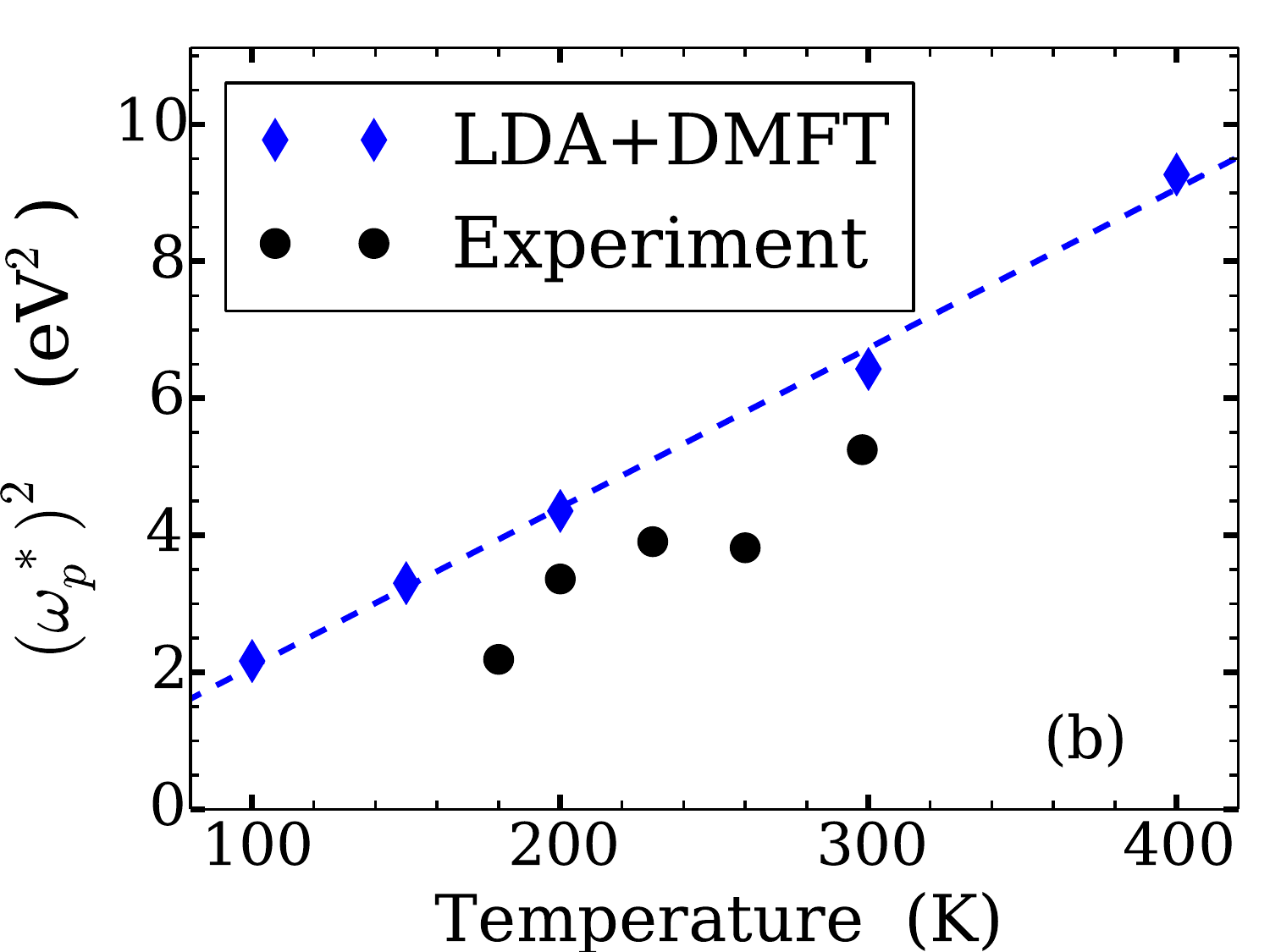}
\includegraphics[width=0.47\columnwidth]{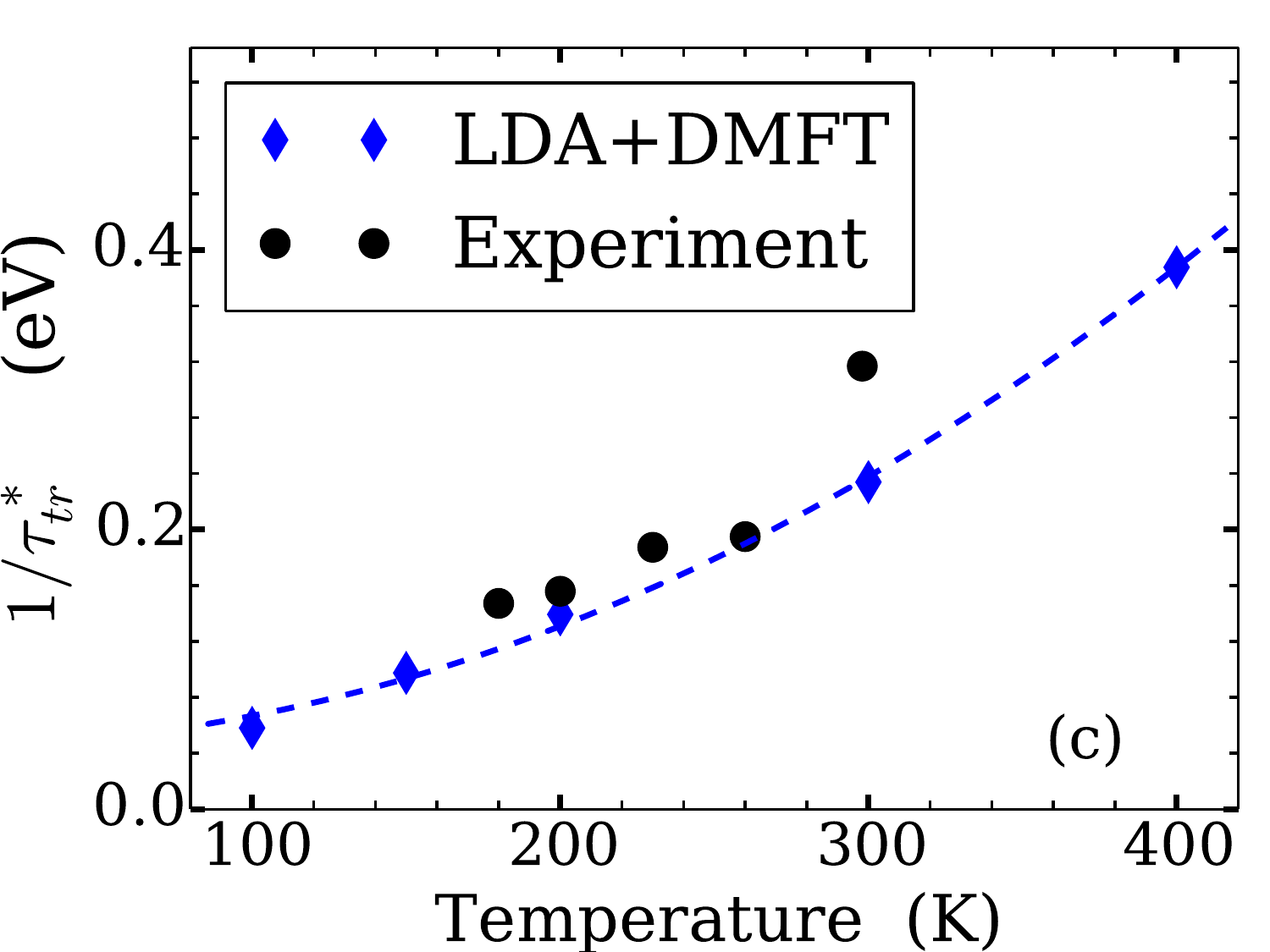}
\caption{ (a) Optical conductivity of \V2O3 calculated with
  LDA+DMFT. The effective plasma frequency (b) and effective
  scattering rate (c) are extracted using
  Eqn.~\ref{eq:lowfrequencyDrude} and compared to those extracted from
  experimental data. Dashed lines are guides to the eyes by fitting
  $(\omega_p^*)^2$ and $1/\tau_{tr}^*$ to linear and parabolic
  functions respectively.}
\label{DMFTOPTIC_V2O3} 
\end{figure}

We calculate the optical conductivity in a broad temperature range as
shown in Fig.~\ref{DMFTOPTIC_V2O3}(a). The main feature of the
experimental optical conductivity, the Drude peak and the shoulder
structure at about $0.1\eV$ as well as their scale , are reasonably
reproduced in our calculations. The Drude peak is gradually diminished
and merges with the shoulder structure at around $400\K$, in agreement
with experiments \cite{V2O3-Optics-Baldassarre}. Therefore LDA+DMFT
calculation provides a satisfactory description of the optical
properties of \V2O3. From the optical conductivity $(\omega_p^*)^2$
and $1/\tau_{tr}^*$ are extracted using
Eqn.~\ref{eq:lowfrequencyDrude}. As shown in
Fig~\ref{DMFTOPTIC_V2O3}(b)(c), they agree reasonably well with those
extracted from experimental data.  In particular the same trends found
with the experimental data, thus the main charactersistics of the
"hidden" Fermi liquid behavior, show up more clearly in the broad
temperature range studied in our calculations: $(\omega_p^*)^2$
appears linear and a $1/\tau_{tr}^*$ appears quadratic versus
temperature. Therefore the proposed analysis of both the experimental
data and the first principle calculations reveals significant
temperature dependence of QPs in terms of $(\omega_p^*)^2$ and an
extended quadratic temperature dependence of $1/\tau_{tr}^*$, but not
of $1/\tau(0)$ .

To further understand the observations, let us recall the QP
interpretation of low frequency optical conductivity in  the DMFT treatment of the doped single
band Hubbard model. In this case,   %  
%\begin{equation}
$\sigma(\omega)|_{\omega\rightarrow 0}=2
Z_{qp}\Phi^{xx}(\bar{\mu})\frac{1}{-i\omega+2/\tau_{qp}^*}
$,
%\end{equation}
in which $Z_{qp}$ and $\tau_{qp}^*$ are the QP weight and life time,
$\Phi$ is the transport function
$\Phi^{xx}(\epsilon)=\sum_{\mathbf{k}}(\partial
\epsilon_{\mathbf{k}}/\partial
\mathbf{k}_x)^2\delta(\epsilon-\epsilon_{\mathbf{k}})$ and $\bar{\mu}$
is the effective chemical potential of QPs\cite{Model-HiddenFL-Xu}. Applying the analysis in
Eqn.\,\ref{eq:lowfrequencyDrude}, we have $(\omega_p^*)^2=8\pi
Z_{qp}\Phi^{xx}(\bar{\mu})$ and $1/\tau^*_{tr}=2/\tau_{qp}^*$,
therefore in this simple model, $(\omega_p^*)^2$ and $1/\tau^*_{tr}$
directly relate to the QP weight and life time. In situations where
$\Phi(\bar{\mu})$ varies little with temperature, the observations
above imply a strong temperature dependence of $Z$ and
$1/\tau_{qp}^*$. We emphasize that although a strong dependence of
scattering rate is generally expected in a Fermi liquid, the
temperature dependence of QP weight is not, but it was observed in
model studies\cite{Model-RQP-Deng, Model-HiddenFL-Xu}.

\begin{figure}
\includegraphics[width=0.47\columnwidth]{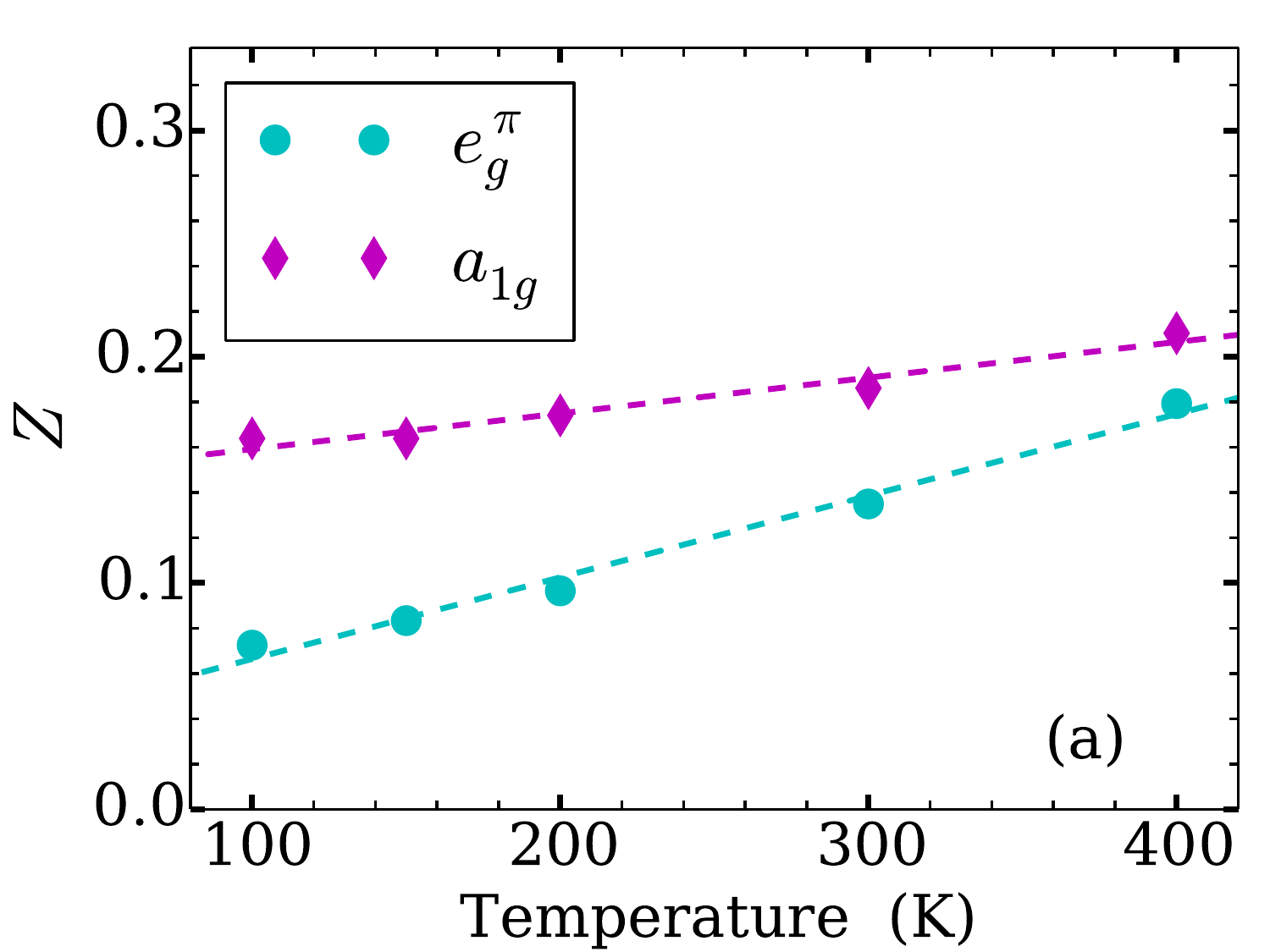}
\includegraphics[width=0.47\columnwidth]{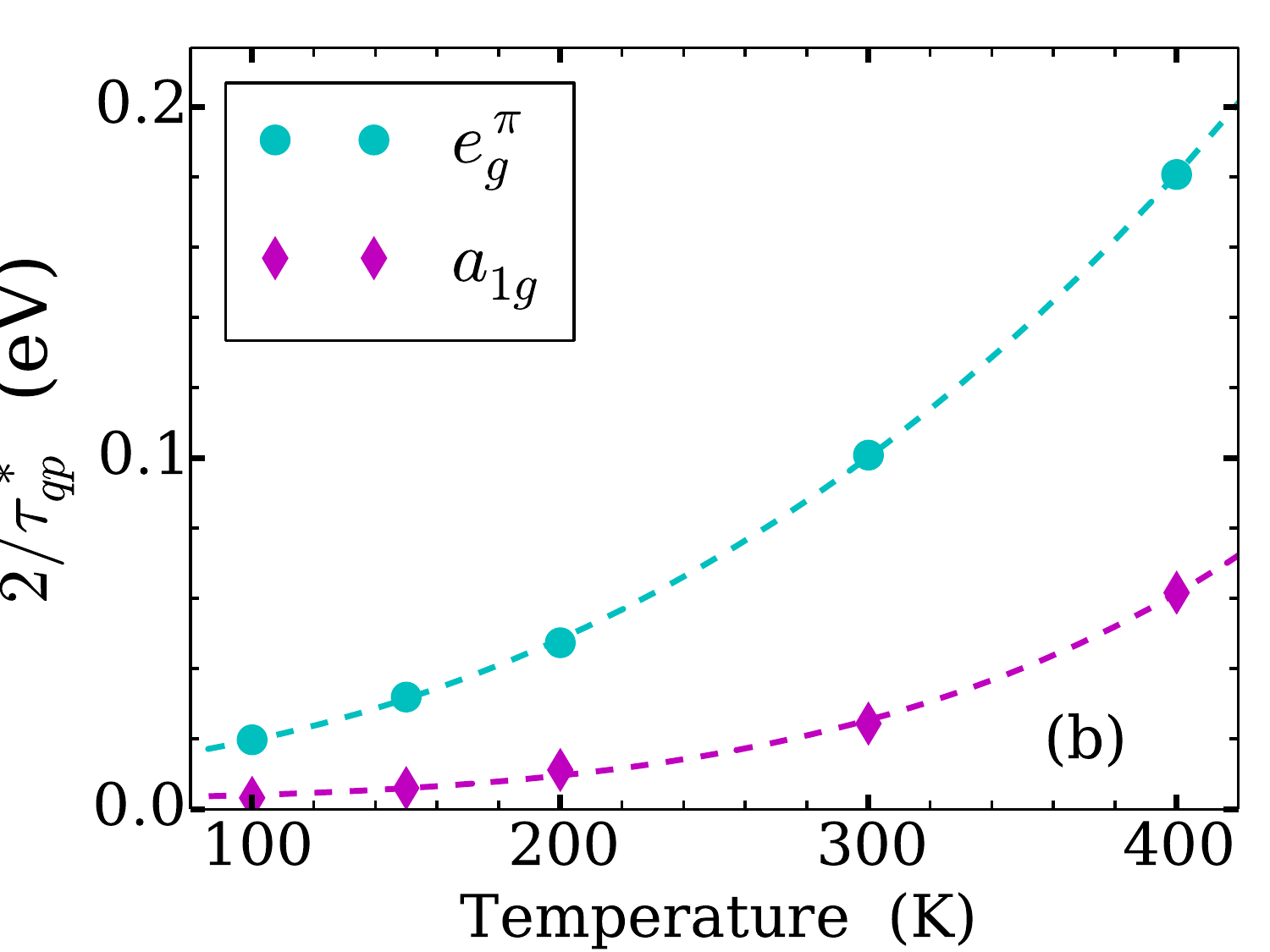}
\caption{ The temperature dependence of (a) QP weight
  $Z=(1-\frac{\partial Re\Sigma(\omega)}{\partial \omega})^{-1}$ and
  (b) effective scattering rate $2/\tau^*_{qp}=-2ZIm\Sigma(0)$ of
  \V2O3 extracted from LDA+DMFT self energies for $e_g^{\pi}$ and
  $a_{1g}$ orbitals. Dashed lines are guides for the eyes by fitting
  $Z$ and $2/\tau^*_{qp}$ to linear and parabolic functions
  respectively}
\label{DMFT_V2O3} 
\end{figure}

We then extract from our calculated self energies, the QP weight
(which is the inverse of the mass enhancement within single site DMFT
) defined as $Z=(1-\frac{\partial Re\Sigma(\omega)}{\partial
  \omega})^{-1}= m_b/m^*$ and the scattering rate defined as
$2/\tau_{qp}^*=-2ZIm\Sigma(0)$.  The QP weight and the QP scattering
rate are shown in Fig.~\ref{DMFT_V2O3}(a)(b).  There is orbital
differentiation between  $e_g^{\pi}$ and $a_{1g}$orbitals as
pointed out in earlier LDA+DMFT studies \cite{V2O3-DMFT-Held,
  V2O3-DMFT-Poteryaev}. Both orbital self energies exhibit the same
trends observed in the studies of model hamiltonians: $Z$ increases
almost linearly with temperature and $1/\tau_{qp}^*$ is nearly
quadratic in temperature for each orbital in the temperature range
considered. Note that $e_g^{\pi}$ orbitals have a much larger spectra
weight at the Fermi level than the $a_{1g}$ orbital thus dominate the
transport. This pronounced temperature dependence is consistent with
that of $(\omega_p^*)^2$ and $1/\tau_{tr}^*$ extracted from optical
conductivity. Therefore the properties of the underlying QPs,
especially the temperature dependence of the QP weight and the QP
scattering rate, are captured in our analysis on optical
conductivities.  We note that in addition the temperature dependence
of underlying QPs manifest itself in the temperature dependence of the
effective chemical potentials of QPs (see online supplementary), which
also contribute to the the temperature dependence of $(\omega_p^*)^2$.

\begin{figure}
\includegraphics[width=0.47\columnwidth]{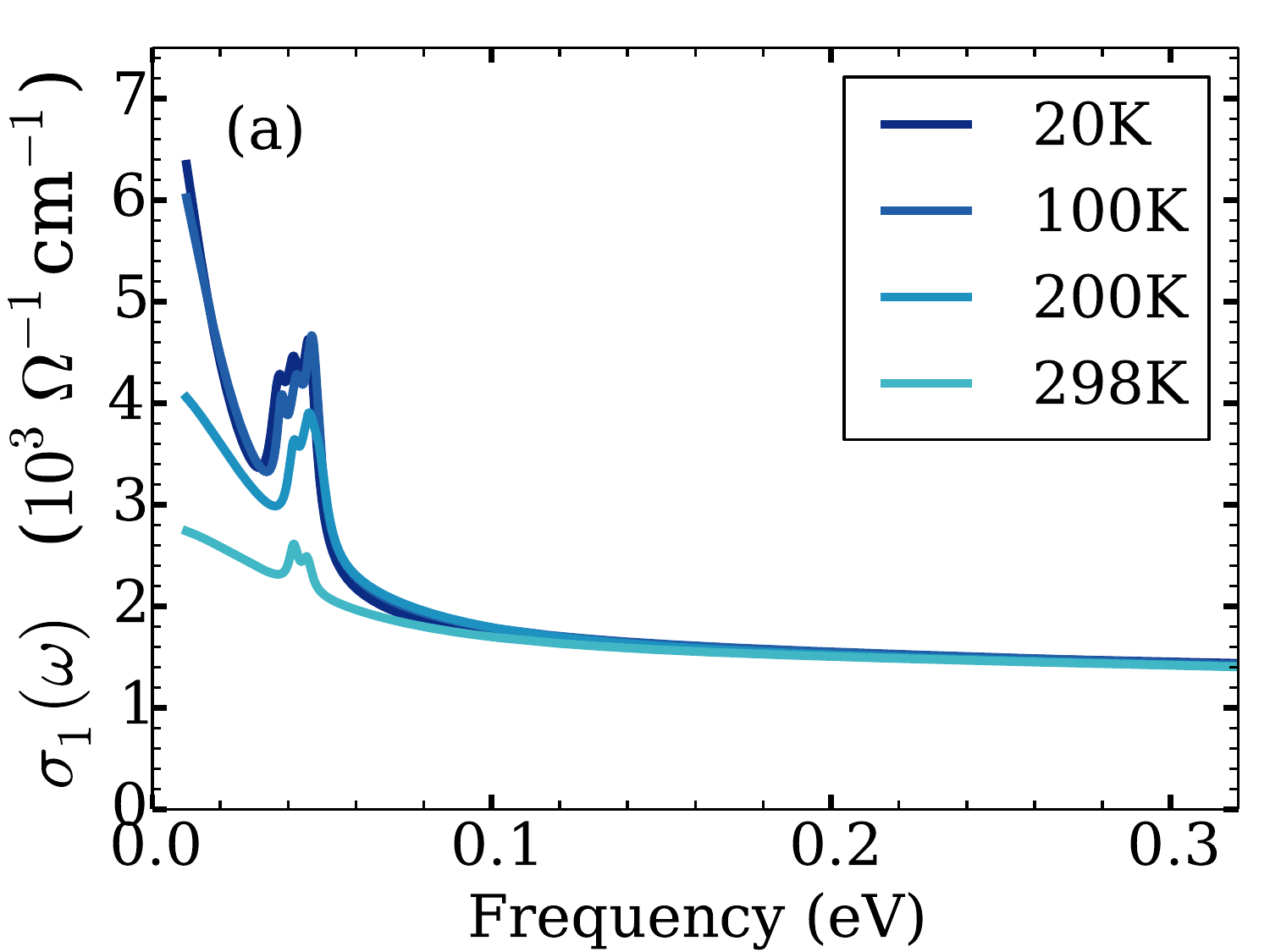}
\includegraphics[width=0.47\columnwidth]{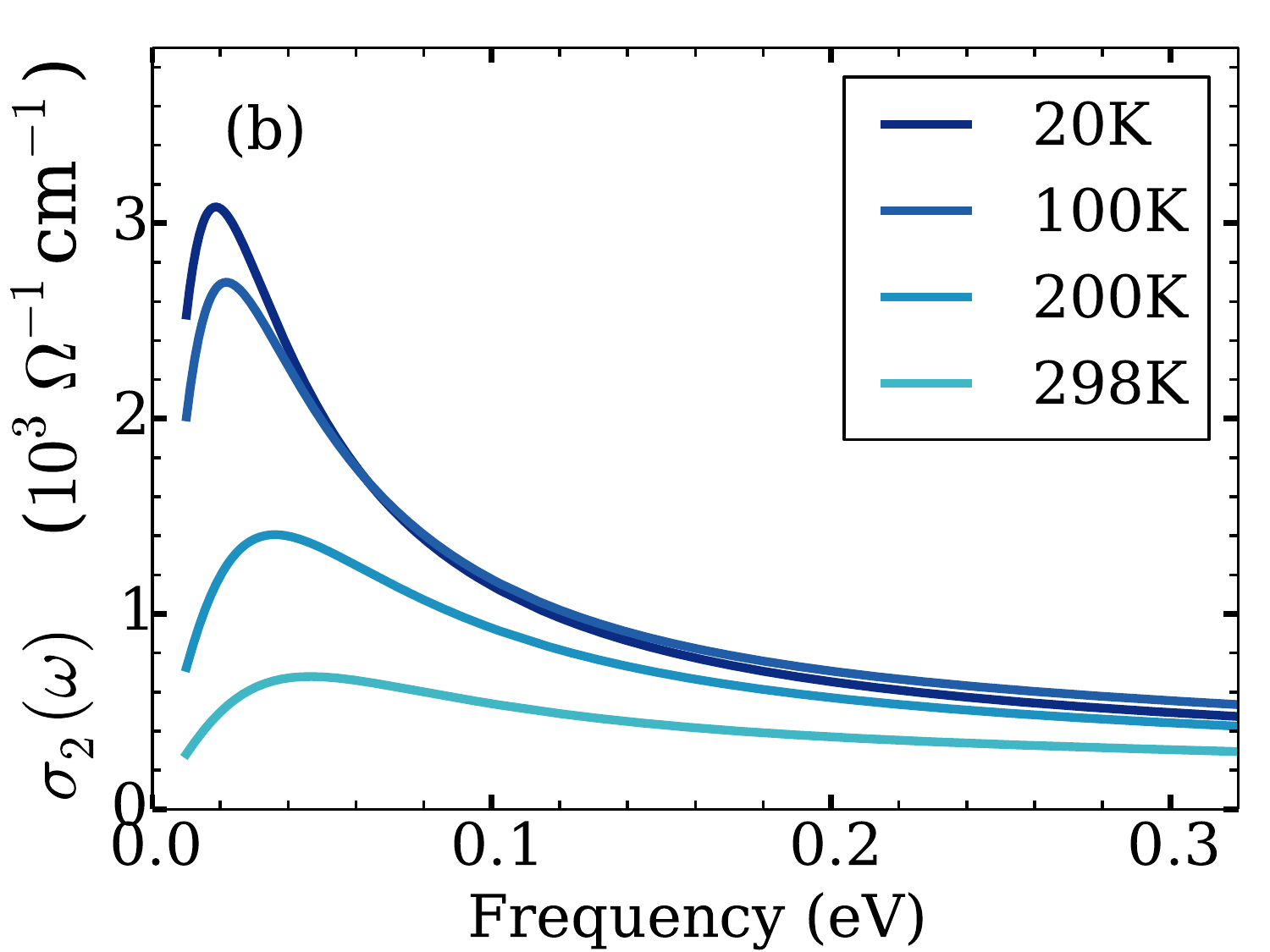}
\includegraphics[width=0.47\columnwidth]{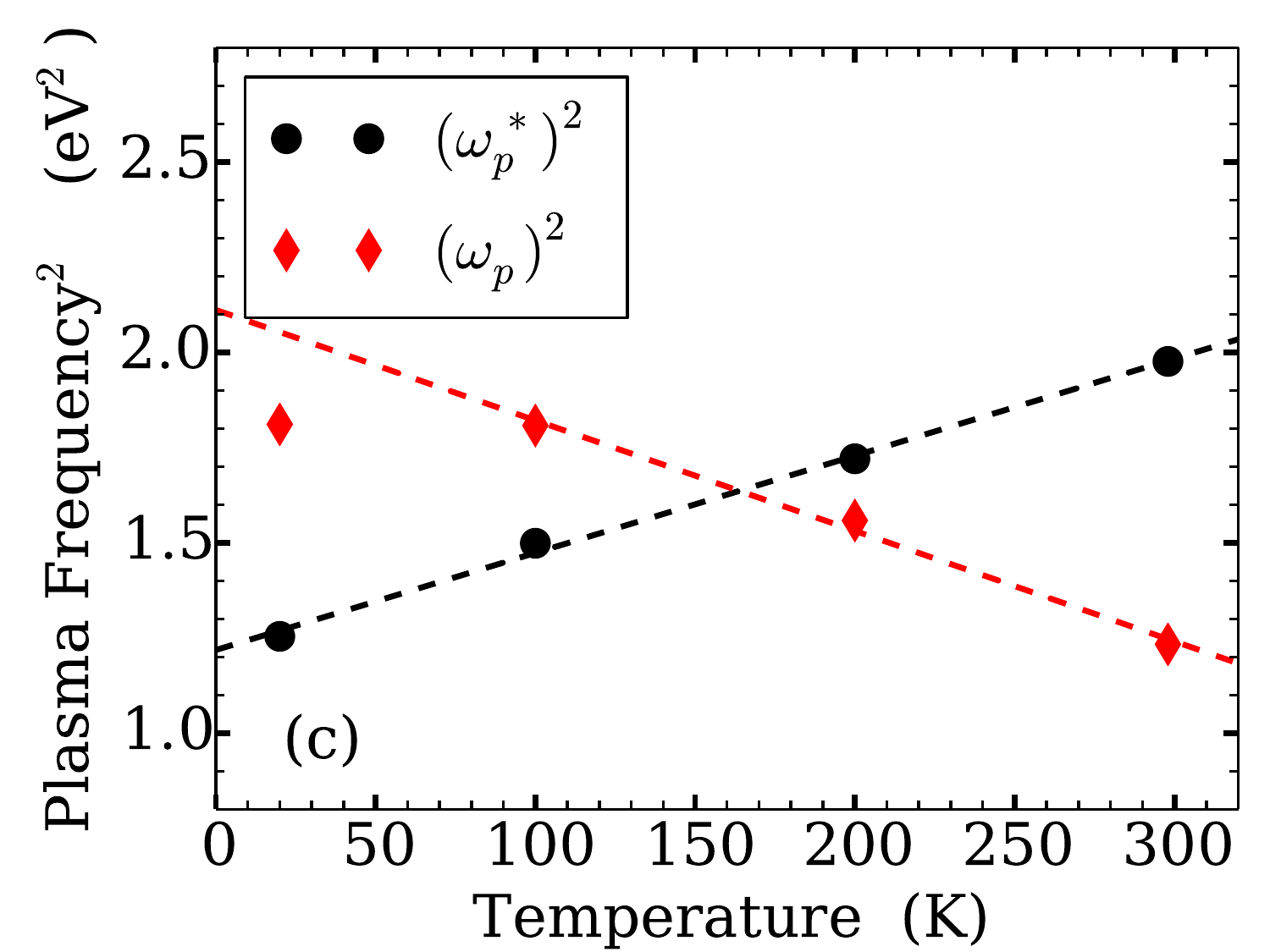}
\includegraphics[width=0.47\columnwidth]{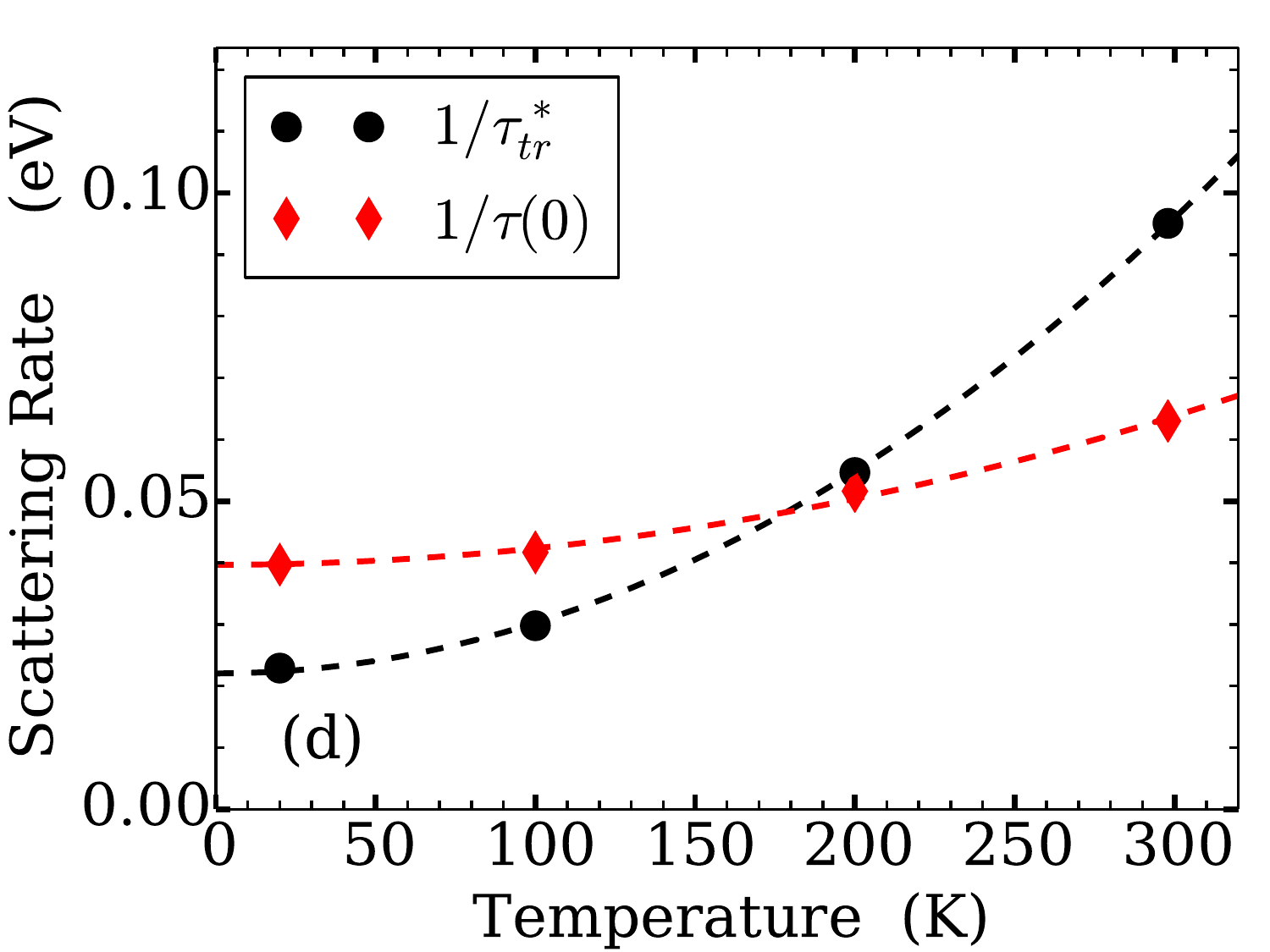}
\caption{Optical conductivity (a) $\sigma_1(\omega)$ and (b)
  $\sigma_2(\omega)$ of NNO film on LAO substrate at different
  temperature is taken from Ref.~\cite{NNO-Optics-Stewart}, from which
  (c) $(\omega^*_p)^2$ and (d) $1/\tau_{tr}^*$  are extracted
  according to Eqn.\,\ref{eq:lowfrequencyDrude}. For comparison,
  $\omega_p^2$ and $1/\tau(0)$ extracted from the extended Drude
  analysis are shown. Dashed lines are guides for the eyes by fitting
  $(\omega_p^*)^2$ ($\omega_p^2$ excluding $T=20\K$) and
  $1/\tau_{tr}^*$ ($1/\tau(0))$ to linear and parabolic functions
  respectively.}
\label{wsq_NNO} 
\end{figure}

We expect that this picture of anomalous transport in correlated
materials 
%\ist{arising from temperature dependence of the electronic
%  structure, and consequently $(\omega_p^*)^2$} 
is not limited to V2O3
and is in fact generally applicable to various strongly correlated
metals .  To check the validity of this general conjecture we apply
the same analysis to experimental data of \NNO (NNO) film on \LAO
(LAO) substrate. NNO is another typical correlated material exhibiting
temperature-driven MIT\cite{Phase-Nickelates-Torrance}. While deposited as film on LAO
substrate, the MIT can be quenched so that it remains metallic down to
very low temperature \cite{NNOfilm-Res-Liu}. High quality optical
conductivities of NNO film are taken from
Ref\;~\onlinecite{NNO-Optics-Stewart} as shown in
Fig~\ref{wsq_NNO}(a)(b). We note that the resistivity is not
$T^2$-like except possibly at the lowest temperature $T=20K$
\cite{NNOfilm-Res-Liu}. We perform the same analysis as above in
\V2O3. $(\omega_p^*)^2$ and $1/\tau_{tr}^*$ are shown in
Fig.~{\ref{wsq_NNO}}(c)(d) in comparison with $\omega_p^2$ and
$1/\tau(0)$ obtained by extended Drude analysis with a cutoff of
$\Omega=125\meV$. Again we have the same features as in \V2O3:
$(\omega_p^*)^2$ is linear in temperature and has the opposite trend
with $(\omega_p)^2$, while $1/\tau_{tr}^*$ has a more pronounced
quadratic behavior in a wide temperature range well above $T_{LFL}$.

In conclusion, in this Letter we point out that the anomalous
transport properties observed in many transition metal oxides arise
from a temperature dependent $(\omega_p^*)^2$ and $1/\tau_{tr}^*$. We
establish that by analyzing both the experimental and the theoretical
data.  This scenario calls for further investigations in other
compounds, starting from systems where there are already preliminary
indications that it applies, for example, CaRu$O_3$ where a similar
low energy analysis was performed \cite{CaRuO3-Drude-Kamal}.  In many
other systems, such as nickelate and pnictides, a temperature
dependent $m^*(0)/m_b$ is seen in the extended Drude analysis
\cite{LNO-Optics-Stewart, Pnictides-Optic-Qazilbash}. Thus an
extraction of the low energy effective plasma frequency
$(\omega_p^*)^2$ as outlined in this paper, and a comparison with
$(\omega_p)^2$ from the partial sum rule would be illuminating.
Recent optical spectroscopy studies of underdoped cuprates
\cite{Cuprates-FL-Mirzaei} at low temperatures revealed temperature
independent $(\omega_p^*)^2$ and a quadratic temperature dependence in
$1/\tau^*_{tr} $, consistent with the earlier theoretical predictions
of cluster DMFT \cite{Superconductivity-CDMFT-Kristjan}. It would be
interesting to extend the measurements to higher temperatures where
deviations from canonical Fermi liquid are expected.  Finally high
resolution studies using spectroscopies such as ARPES and STM in
quasiparticle interference mode would be very useful to separate the
various contributions to the temperature dependence of
$(\omega_p^*)^2$ by probing directly the electronic structure.

We acknowledge very useful discussions with A. Georges and P. Armitage.  This work was
supported by NSF DMR-1308141 (X. D. and G.K ), NSF DMR 0746395 (K. H.)
and DOE-BES ( A. S. and D. B.).

\appendix
\begin{widetext}
\section{temperature dependence of momentum-resolved spectra and quasiparticle bands of \V2O3}
Electronic structure of correlated metal has a significant
temperature dependence. This can be seen in the momentum-resolved
spectra, defined as
\begin{equation} 
A(\mathbf{k},\omega)=-\frac{1}{\pi}
Im[\frac{1}{\omega+\mu(T)-H_{\mathbf{k}}-\Sigma_\mathbf{k}(\omega,T)}
\label{QP}
\end{equation}
 in LDA+DMFT calculations, in which $H=-\nabla^2
+V_{ext}+V_{H}+V_{xc}-\hat{E}V_{dc}$,
$\Sigma_\mathbf{k}(\omega,T)=\hat{E}_{\mathbf{k}}\Sigma(\omega,T)$,
$\hat{E}$ is the embedding operator and $\Sigma(\omega,T)$ is
the impurity self energy in orbital space. As mentioned in the main text, the
quasiparticles (QPs) are well defined and meaningful for the transport
properties even when the scattering rate in the self energy is
large. The QP dispersion $\epsilon_{\mathbf{k}}^*$ is defined as the
solution to the following equation:
\begin{equation}
\det (\omega+\mu(T)-H_{\mathbf{k}}-Re[\Sigma_{\mathbf{k}}(\omega,T)])=0.
\label{QP}
\end{equation}

The spectra $A(\mathbf{k},\omega)$ and the QP dispersion
$\epsilon^*_{\mathbf{k}}$ for two different temperature are depicted
in Fig.~{\ref{Spectra_DMFT_V2O3}}(a)(b). We find that increasing
temperature broadens the spectra $A(\mathbf{k},\omega)$ significantly
due to the temperature dependence of QP scattering rate. The
temperature also modifies the QP dispersion in two aspects: with
increasing temperature the QP bands near the Fermi level are less
renormalized in accordance with the temperature dependence of QP
weight shown in the main text, and the QP bands are shifted
accordingly. Both effects affect the transport properties.  The shifts
of the quasiparticles can be described with the effective chemical
potential $\bar{\mu}(T)=\mu-Re\Sigma(0,T)$, which in general has an
orbital index due to orbital
differentiation. Fig.~{\ref{Spectra_DMFT_V2O3}}(c) shows the
temperature dependence of $\bar{\mu}(T)_{e_g^{\pi}}$ and
$\bar{\mu}(T)_{a_{1g}}$. The shifts of QP bands result in a change of
Fermi surface upon increasing temperature.  We note that our predicted
QP bands with full electron calculations are different from those
calculated with a downfolded Hubbard model \cite{V2O3-DMFT-Poteryaev},
where the orbital polarization was found to be very large and only a
single Fermi surface remains. This is because the crystal field
splitting of \V2O3 is mainly hybridization driven and thus better
described with $p$-$d$ like model rather than the Hubbard model. This
prediction is to be verified by further detailed experiments
especially with angular-resolved photoemission spectroscopy
measurements.
\begin{figure}[h]
\includegraphics[width=0.32\columnwidth]{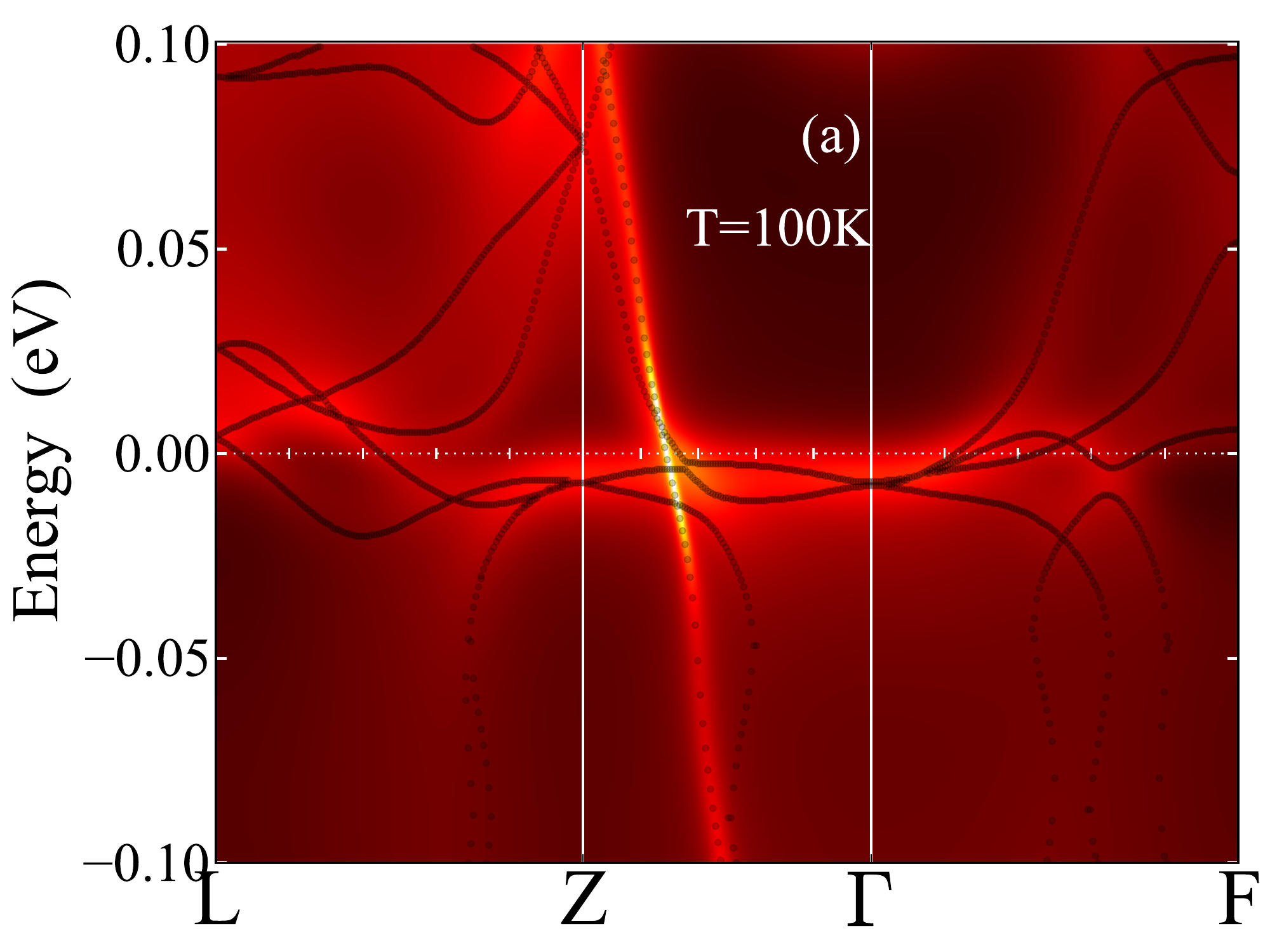}
\includegraphics[width=0.32\columnwidth]{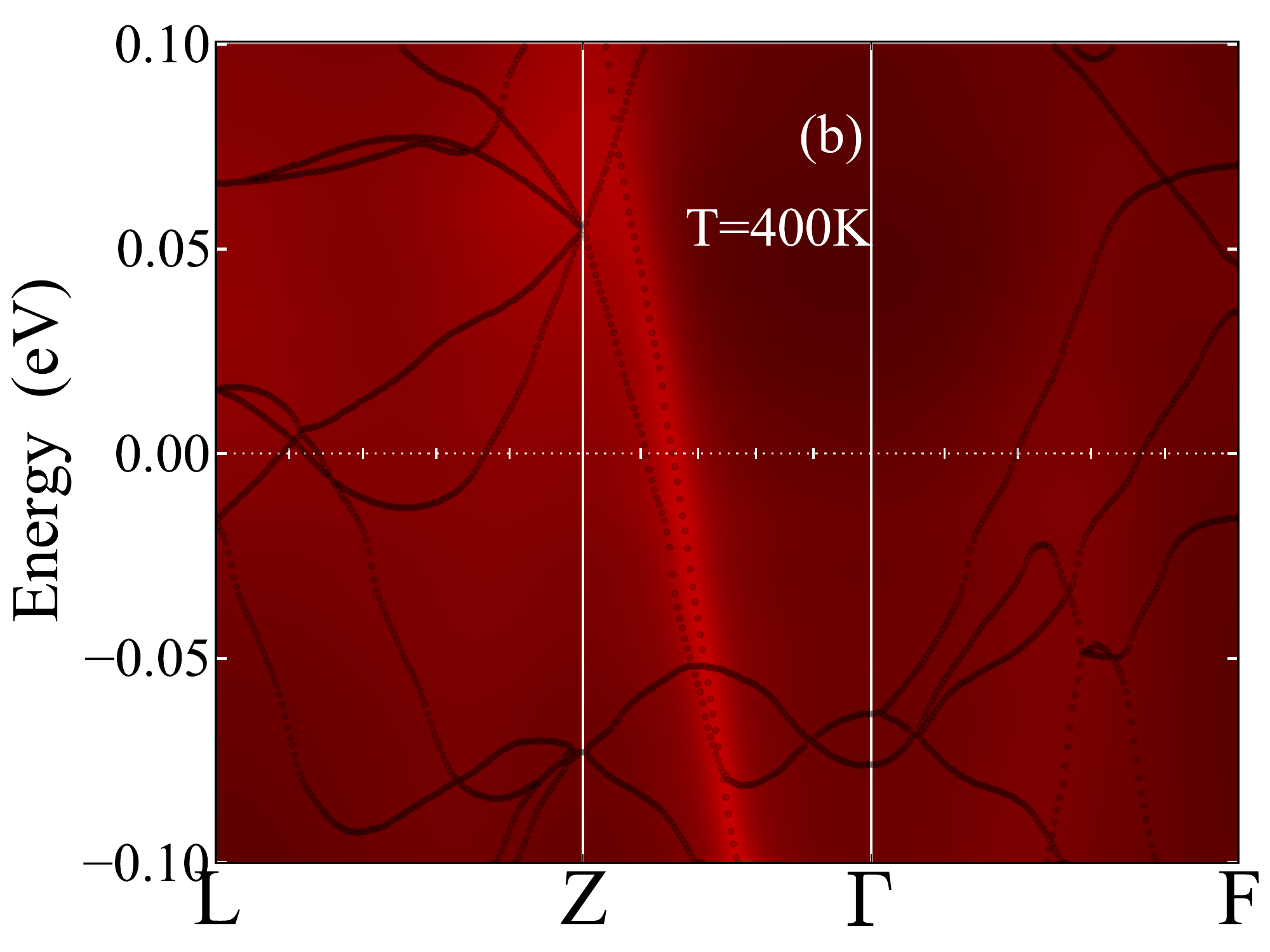}
\includegraphics[width=0.32\columnwidth]{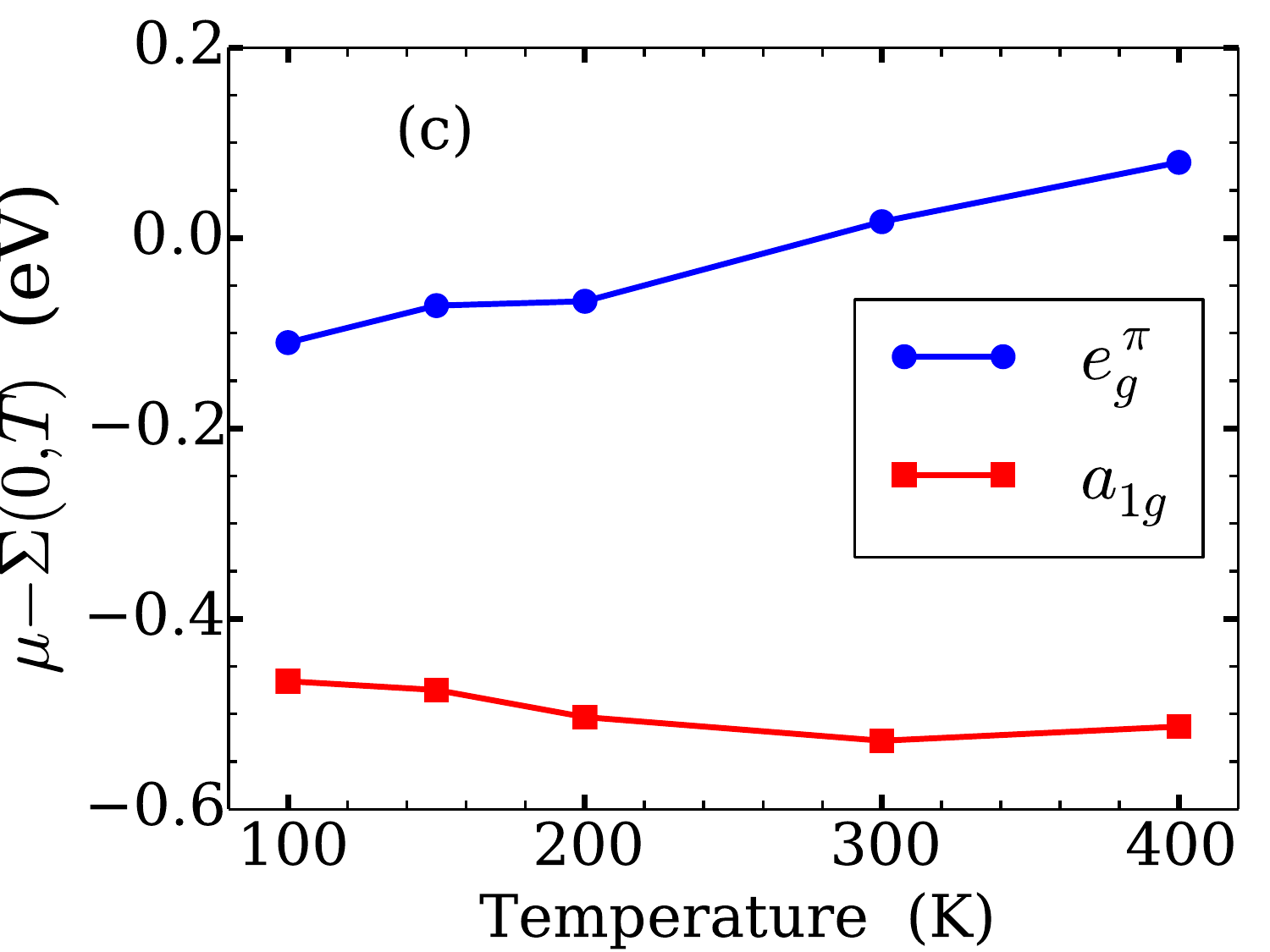}

\caption{The calculated momentum-resolved spectra of \V2O3 at
  temperature $T=100\K$ (a) and $T=400\K$ (b). The corresponding QP
  dispersion $\epsilon^*_{\mathbf{k}}$ computed using Eqn.\ref{QP} is
  shown with dots.  The effective chemical potentials for $e_g^{\pi}$
  and $a_{1g}$ orbitals are shown in (c).  }
\label{Spectra_DMFT_V2O3} 
\end{figure}

\end{widetext}

%\bibliography{refs}

\end{document}